# Electric-field induced droplet vertical vibration and horizontal motion: Experiments and numerical simulations

Ziqi Li[1†], Yongzhou Luo[1†], Rucheng Dai[2*], Zhongping Wang[2*] and Xiaoyu Sun[2]

[1]*School of Physical Sciences,*
*University of Science and Technology of China, Hefei, Anhui 230026, P.R. China*
[2]*Centre of Physical Experiments,*
*University of Science and Technology of China, Hefei, Anhui 230026, P.R. China*
*Corresponding authors: dairc@ustc.edu.cn; zpwang@ustc.edu.cn

Author Contributions
[†]Z.Q. Li & Y.Z. Luo contributed equally to this work

**Abstract**

In this work, Electrowetting on Dielectric (EWOD) and electrostatic induction (ESI) are employed to manipulate droplet on the PDMS-ITO substrate. Firstly, we report large vertical vibrations of the droplet, induced by EWOD, within a voltage range of 40 to 260 V. The droplet's transition from a vibrating state to a static equilibrium state are investigated in detail. It is indicated that the contact angle changes synchronously with voltage during the vibration. The electric signal in the circuit is measured to analyze the vibration state that varies with time. By studying the influence of driving voltage on the contact angle and the amplitude in the vibration, it is shown that the saturation voltage of both contact angle and amplitude is about 120 V. The intrinsic connection between contact angle saturation and amplitude saturation is clarified by studying the surface energy of the droplet. A theoretical model is constructed to numerically simulate the vibration morphology and amplitude of the droplet. Secondly, we realize the horizontal motion of droplets by ESI at the voltage less than 1000 V. The charge and electric force on the droplet are numerically calculated. The frictional resistance coefficients of the droplet are determined by the deceleration of the droplet. Under consideration of frictional resistance of the substrate and viscous resistance of the liquid, the motion of the droplet is calculated at 400 V and 1000 V, respectively. This work introduces a new method for manipulating various forms of droplet motion using the single apparatus.

# 1. Introduction

Microfluidics refers to a new type of manipulation technology that independently controls discrete droplets, which has important application potential in liquid lens, chemical synthesis, biochemical analysis and so on.[1-3] Electric field, magnetic field, and sound waves have been employed to manipulate the droplets.[4-6] The electric field microfluidic technology includes electrowetting on dielectric (EWOD),[7,8] light-induced electric field induction[9,10] and electrostatic induction (ESI).[11-13] In 1875, Lippmann pioneeringly proposed the Lippmann equation to describe the phenomenon of electrowetting during the electrocapillary investigation.[14] In 1993, Berge established the Lippmann-Young equation based on the investigations from Lippmann and Yang, and opened a new field of quantitative study of EWOD.[15] Recently, a new electrowetting mechanism that using electric field induced attachment and detachment of ionic surfactants to the substrate is reported.[16] In terms of practical application, digital microfluidics based on the most mature control technology of EWOD enables programmable manipulation of droplet movement, distribution, splitting and merging through microelectrode arrays.[17-19] Ahmad et al.[20] found the maximum change of 28º for the contact angle of the droplet from the oscillation phenomenon of droplets under the sine wave electric field in EWOD. Quintero et al.[21] pointed out that the larger the signal frequency, the larger the amplitude and the longer the oscillation time for the droplet oscillation induced by a single sinusoidal signal in EWOD. Lu et al.[22] used the MKT-based method to simulate the morphological changes of droplets driven by direct current (DC) and alternating current (AC). They found that the droplet's spreading and contracting time at DC model can be used as characteristics of the droplet's external voltage response time. At AC model, the higher the vibration of the mode, the smaller the amplitude of the droplet. Chakraborty et al.[23] studied the influence of the delay time and voltage of the square wave pulse in EWOD on the vibration of the droplet, pointing out that the vibration frequency decreases with the increase of delay time, while the amplitude increases with the increase of voltage. Tröls et al.[24] used Finite Element Method (FEM) to simulate the droplet vertex position with time before and after DC application, predicted the existence of the oscillation behavior of DC driving droplet. Dwivedi et al.[25] explored the spreading process and the morphological changes of droplets under EWOD by COMSOL-Multiphysics simulation. The results show that the contact angle decreases with the movement of the contact perimeter, and after the saturation point, the droplets oscillate and gradually tend to equilibrium. Hong et al.[26] studied the droplet bounce under square wave pulse drive, compared with DC constant voltage drive, the threshold voltage required for droplet bounce is reduced by 70% (from 100 V to 30 V), and the droplet bounce height is increased by 50%. Liu et al.[27] studied the energy conversion of the droplet bouncing process triggered by EWOD, and pointed out that the mutual conversion of surface energy, kinetic energy, gravitational potential energy and internal energy occurs during the droplet bouncing process. The peak conversion efficiency of surface energy to gravitational potential energy can reach more than 30%. Lee et al.[28] studied the effect of different electrode configurations on the bounce height of droplets, and obtained that the bounce height of the vertical needle tip electrode was 2.5 mm, the bounce height of the surface needle tip electrode was 2.2 mm, and the bounce height of the buried electrode was 2.7 mm. Another new droplet manipulation technology is electrostatic induction method, which overcomes the shortcomings of digital microfluidic technology that droplets can only move along a fixed track in one dimension, and realizes the free control of droplet movement in a two-dimensional plane. Jin et al.[11] reported an electrostatic tweezer (DEST) droplet manipulation technique to achieve non-contact manipulation of droplets at thousands of volts, and simulated the amount of charge and charge distribution of droplets. Dai et al.[12] developed a high-speed and controllable droplet manipulation platform, which uses the principle of electrostatic induction to achieve hundreds of millimeters per second's movement and rapid stop of droplets. Pavliuk et al.[13] proposed a multifunctional control platform based on buried electrodes to achieve fast control of droplets. However, thousands of volts significantly limit the application of electrostatic induction manipulation technology. It is urgent to develop a type of low volts manipulating technique for the wide application of microfluidic. Besides, in terms of the theory of electrostatic induction control technology, the existing work has only carried out theoretical analysis of the charge distribution and charge quantity of droplets, and has not carried out theoretical simulation research on droplet motion under electrostatic field.

In this paper, the large vertical vibration and horizontal motion of droplets induced by electric field in a low-resistance environment was reported. First, The Polydimethylsiloxane (PDMS) films were coated on the surface of ITO glass (PDMS-ITO), the tetrapropoxysilane (TPOS) was used as the surrounding environment of the droplet. Second, the large vertical vibration of the droplet was caused by EWOD. With the characteristics of large amplitude and no attenuation, the vibration we report is significantly different from the EWOD induced oscillation phenomenon that reported previously.[26,27] Then, the horizontal motion of droplets induced by ESI was achieved at a voltage below 1000 V. The motion of the droplet was quantitatively calculated and the results showed a good agreement with the experiments. The quantitative measurement and simulation of the horizontal motion of the droplet induced by ESI were not seen in the previous works to we authors' knowledge.

**2.Research Methods**

**2.1 Experiment Method**

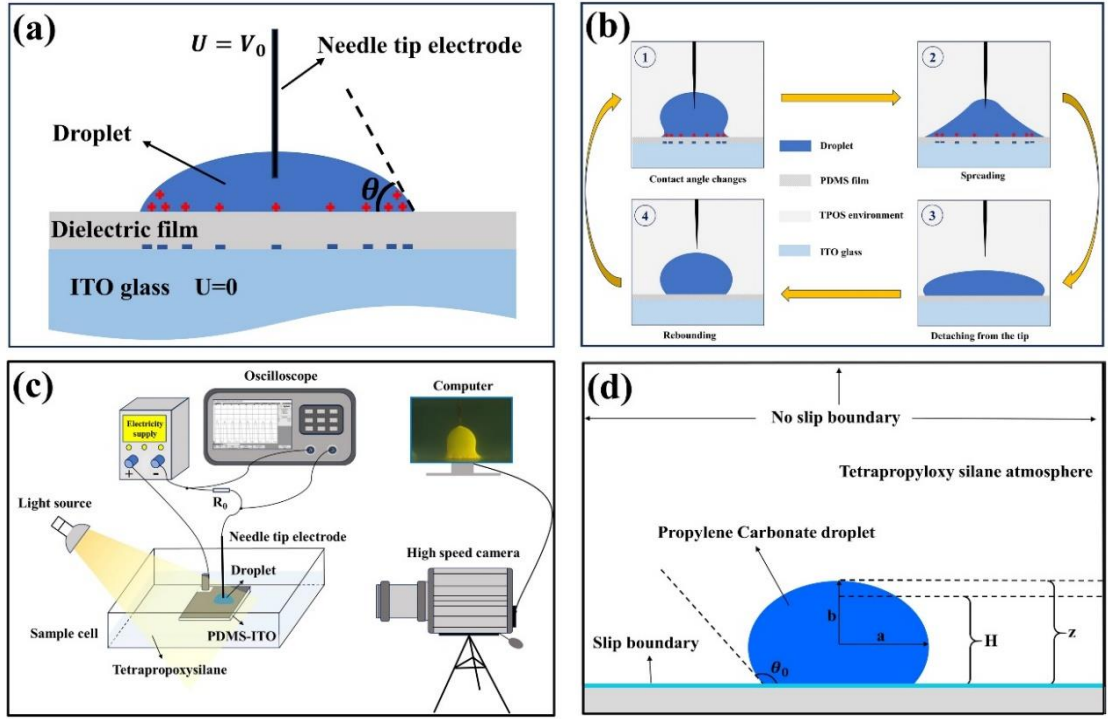

FIG. 1. (a) Schematic diagram of the ordinary EWOD; (b) The cyclic process of the vibration; (c)Schematic diagram of the experimental setup; (d) Schematic diagram of the initial shape of the droplet (the length of semi-major axis is a, the length of semi-minor axis is b, the distance from the apex of the droplet to the ITO substrate is z , the equilibrium contact angle is $\theta_0$, the height of the needle tip electrode from the ITO substrate is H, and the distance of the droplet to the boundary is much larger than the droplet size).

With a dielectric layer added between the substrate electrode and the droplet, the EWOD refers to the phenomenon that when an electric field is introduced between the droplet and the PDMS-ITO substrate the wettability between the droplet and the substrate increases, and the contact angle decreases, thus the overall shape of droplet will change with electric field. The Schematic diagram of the ordinary EWOD is shown in Fig.1(a). When the needle tip electrode is inserted into the droplet and the droplet reaches the static equilibrium state, the contact angle of droplet and the applied voltage satisfy the Lippmann-Young equation:[15]

$$\cos\theta = \cos\theta_0 + \frac{c}{2\gamma_{LL'}}U^2 \quad (1)$$

where $\theta$ is the contact angle, $U$ is the driving voltage, $\theta_0$ is the intrinsic contact angle of the droplet in the liquid-phase environment, $\gamma_{LL'}$ is the surface tension coefficient between the droplet and the surrounding environment, and $c$ is the capacitance per unit area of the PDMS. In this paper, the large vertical vibration of the droplet on the surface of the PDMS film is a novel physical phenomenon based on EWOD. When the needle tip electrode contacts the droplet, the charge is conducted to the surface of the droplet, causing a change in the surface tension between the droplet and the substrate, which in turn leads to a decrease in the droplet's contact angle. Then the overall height of the droplet decreases and detaches from the needle tip electrode, the charge on the droplet is released through the PDMS film, and the contact angle of the droplet is rapidly restored. Under the effect of surface tension, the droplet rebounds upward and contacts the needle tip electrode again, which causes a continuous vibration phenomenon of the droplet. Figure 1(b) shows the cyclic process of the vibration. The ESI is to use the metal electrode to generate an electric field to make the grounded droplet electrically charged, and then manipulate the droplet movement based on electric field force. In this text, under the function of the needle tip electrode, the electric charge is conducted to the droplet through the PDMS films, and the charged droplet is attracted for accelerated motion. Specifically, the PDMS-ITO substrate was submerged in a low-resistance tetrapropoxysilane environment. A high-speed camera was used to capture the vertical vibration and horizontal motion of the droplet. An oscilloscope was used to measure the electrical signals in the circuit during the droplet vibration process. The experimental setup is shown in Fig.1(c), and the detailed experimental method is shown in Appendix 1.

**2.2 Simulation methods**

The large vertical vibration and horizontal motion of the droplet are simulated by COMSOL-Multiphysics software. For the vertical vibration, it can be simplified as a cylindrical symmetric system, as shown in Fig.1(b). The system is composed of two phases of fluids, the liquid environment and the droplet. Since the topological properties of the two fluids remain unchanged throughout the process, the flow of liquids can be simulated using Moving Mesh method, which accurately describes the evolution of phase interfaces and the effects of surface tension. Assuming both fluids are

incompressible, their evolution is governed by the Navier-Stokes equations:

$$\rho \nabla \cdot \boldsymbol{u} = 0 \tag{2}$$

$$\rho \frac{\partial \boldsymbol{u}}{\partial t} + \rho (\boldsymbol{u} \cdot \nabla) \boldsymbol{u} = \nabla \cdot \left[ -p \mathbf{I} + \mathbf{K} \right] + \boldsymbol{F} + \rho \boldsymbol{g} \tag{3}$$

where,

$$\mathbf{K} = \eta \left( \nabla \boldsymbol{u} + (\nabla \boldsymbol{u})^{\mathrm{T}} \right) \tag{4}$$

$\boldsymbol{u}$ is the velocity field, $p$ is the pressure, $\rho$ is the density, $\eta$ is the dynamic viscosity of the fluid, $\boldsymbol{F}$ is the force acting on a unit volume of fluid (also known as the "volume force"), $\boldsymbol{g}$ is the gravitational acceleration. The cross-section of the droplet in the simulation is elliptical (semi-major axis a, semi-minor axis b), and the distance from the position of the highest point of droplet to the base of the ITO is z (exact lengths from experiments), as shown in Fig.1(b).

In the experiment, when the droplet contacts the electrode during the rebound process, the contact angle of the droplet begins to change. Therefore, in the simulation, the theoretical model is constructed as follows: the position of the highest point of droplet is tracked, and when the highest point of droplet rises to the height of the needle tip, H, the equilibrium contact angle (the contact angle of the droplet when it reaches the equilibrium state of EWOD) instantly shifts to $\theta_1$. When the position of the highest point of the droplet is detached from the tip of the needle, the equilibrium contact angle instantly shifts to $\theta_0$. That is, the relationship between the equilibrium contact angle, $\theta$, and the position of the highest point of the droplet, $z$, is:

$$\theta = \begin{cases} \theta_0, & z < \mathrm{H} \\ \theta_1, & z \geq \mathrm{H} \end{cases} \tag{5}$$

where $\theta_0$ can be taken as the contact angle when the droplet is resting on the substrate, and $\theta_1$ is related to the voltage $U$ applied between the needle tip electrode and the substrate. In the theoretical simulations, under voltage $U$, $\theta_1$ is taken to be the average value of the contact angle of the droplet when it is in contact with the needle tip electrode. When simulating the vibration morphology of the droplets at different voltages, some experimental effects were not considered in the process:

(i) the viscosity resistance between the droplet and the substrate;

(ii) the viscosity resistance between the needle tip electrode and the droplet;

(iii) the droplet deformation caused by the attraction of the strong electric field near the needle tip electrode to the top of the droplet;

(iv) the contact angle transition time of the droplet;

When under ESI, a droplet is accelerated in the horizontal direction. We can use COMSOL-Multiphysics to simulate and calculate the electric field force on the droplet. The COMSOL-Multiphysics takes the entire volume of the droplet as the integration domain and calculates the following volume integral to obtain the electric field force on the droplet:

$$\boldsymbol{F} = \iiint \nabla \cdot \boldsymbol{T} \tag{6}$$

where $\boldsymbol{T}$ is the Maxwell stress tensor, its expression is:

$$\boldsymbol{T} = \begin{pmatrix} \varepsilon_0\varepsilon_r(E_x^2 - \frac{E^2}{2}) & \varepsilon_0\varepsilon_r E_x E_y & \varepsilon_0\varepsilon_r E_x E_z \\ \varepsilon_0\varepsilon_r E_x E_y & \varepsilon_0\varepsilon_r(E_y^2 - \frac{E^2}{2}) & \varepsilon_0\varepsilon_r E_y E_z \\ \varepsilon_0\varepsilon_r E_x E_z & \varepsilon_0\varepsilon_r E_y E_z & \varepsilon_0\varepsilon_r(E_z^2 - \frac{E^2}{2}) \end{pmatrix} \tag{7}$$

The droplet sizes and tip heights used in the simulations are obtained from videos of the experiments.

### 3. Results and Discussion

### 3.1 Vertical vibration

### 3.1.1 Three states of droplet motion

With PDMS film thickness of 3.5 μm, droplet of 4 μL, direct current voltage (DC) of 200 V unchanged and decreasing the height of the needle tip electrode, the droplet on the PDMS film could be divided into three stages. In the stage I (non-contact stage), the needle tip electrode did not contact the droplet (H>2.04 mm), the droplet was hemispherical shape without undergoing any vibration. In stage II (large vertical vibration stage), the needle tip electrode contacts the droplet (1.32 mm<H<2.04 mm), and the droplet vibrates vertically. The droplet's vibrating morphology in a vibration cycle at different needle tip electrode height is shown in Fig. 2.

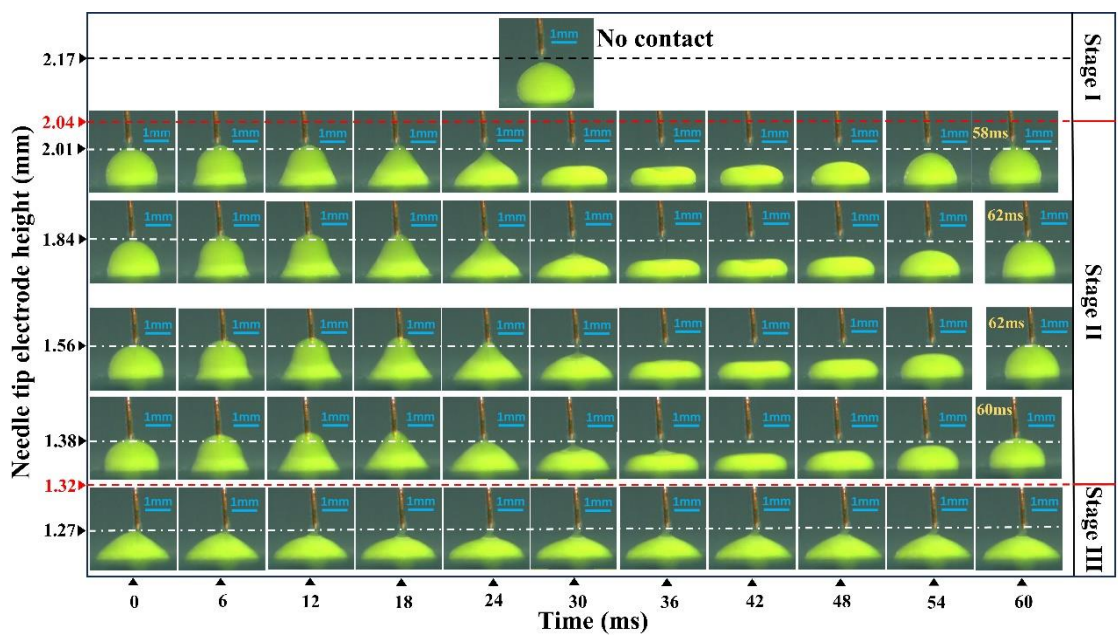

**FIG. 2. The relationship between the state of the droplet and the needle tip electrode height H.**

Taking the images at the needle tip electrode height of 2.01 mm as an example: First 6 ms: the droplet contacts with the needle tip electrode and its contact angle begins to change; Second 24 ms: the droplet gradually evolves into a conical shape, the overall height decreases, and finally detaches from the needle tip electrode. Final 28 ms: the droplet flattens and begins to spring upward, and the droplet evolves into a hemispherical shape and returns to the initial state. As the height of the needle tip electrode was gradually reduced, continuous vibration of the droplet is found and the vibration period is almost unchanged (58 ms when H=2.01 mm, 62 ms when H=1.84 mm and H=1.56 mm, 60 ms when H=1.38 mm). As the height of the needle tip electrode decreases, it can be seen that the morphology evolution of the droplet is similar to the vibration pattern at H=2.01 mm with the amplitude reduced. In the stage III (stable EWOD equilibrium stage), the needle tip electrode inserts into the droplet (H<1.32 mm), the droplet vibration disappears and a stable EWOD equilibrium appears. It is worth noting that in the second stage II, the droplets have a large vibration period of about 60 ms, and the vibration period is almost independent of the needle tip electrode height.

The results show that the large vertical vibration is a non-equilibrium state in the EWOD, which is obviously different from the stable equilibrium state of EWOD. The whole process of the experiment can be seen in video I, "three states of the droplet".

**3.1.2 Contact angle changing**

The contact angles of the droplet dependent on time ($t$) in the vibration state is shown in Fig. 3(a). In the first 6 ms, the droplet rebounds upwards and the contact angle decreases from initial 140° to about 120°. In the successive 4 ms (6~10 ms), keeping the connection between the needle tip electrode and the droplet, and the droplet contact angle is rapidly reduced from 120° to about 40°. In the next 20ms (10~30ms), continuous holding the contacting status, the droplet spreads outward at the bottom and its height decreases, and the contact angle of the droplet is stable at about 40°. In the further 8 ms (30~38 ms), the droplet evolves into a flat shape and detaches from the needle tip electrode; the contact angle increases rapidly from about 40° to about 120°. During 38-60 ms, the droplet rebounds upwards with a contact angle of about 130°. The evolution of shapes for the droplet is shown in Fig. 3(b). Under AC (50 Hz, 200 V), the large vertical vibration of the droplet is induced and its contact angle varies with time is shown in Fig. 3(c) and the evolution of the droplet morphology is shown in Fig.3(d).

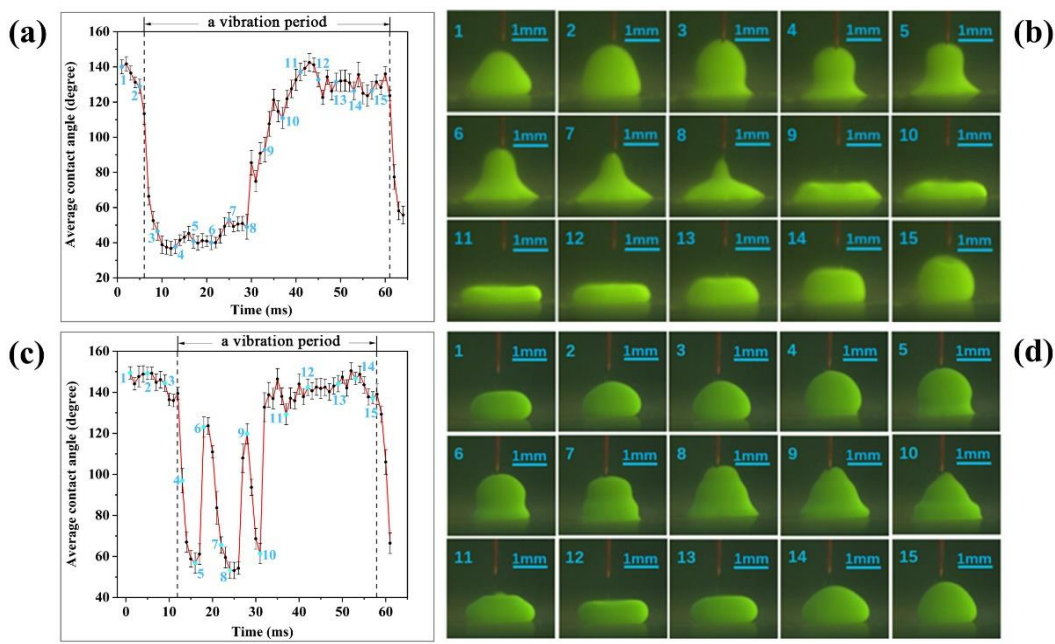

**FIG. 3. Changes in droplet vibration morphology and contact angle at DC and AC voltages**

As shown in the Fig.3(c,d), in the initial 12 ms (0~12 ms), the droplet is free to rebound without contacting the needle tip electrode, and the contact angle of the droplet is about 140°. During 12~32 ms: the droplet is in contact with the needle tip electrode, and due to the alternating voltage, the contact angle of the droplet reaches a minimum value of about 50° at moments 16 ms, 25 ms, and 33 ms, and a maximum value of about 120° at moments 19 ms and 28 ms. During 32~58 ms: the droplet is out of contact with the needle tip electrode, and the droplet rebounds upward with a contact angle of about 140°. It is worth noting that the time span between two adjacent contact angle peaks is 8ms and 9ms respectively, which are very close to half of the alternating current signal cycle. Therefore, it is implied that the contact angle of the droplet is driven by the power supply voltage synchronously. The evolution behaviors of the droplet at AC voltage are significantly different from those at DC voltage.

Figure 4(a) represents the voltage waveform of the sampling resistor $R_0$ (1 kΩ) with time at 200V DC. During a vibration cycle, the droplet contacts with the needle tip electrode, and the voltage signal rises rapidly from 0 to 0.42 V with a rising edge of about 8.5 ms, after which the voltage signal begins to drop and the dropping edge is about 23.0 ms. When the droplet completely detaches from the needle tip electrode, the voltage signal drops to 0. The voltage signal period of the sampling resistor $R_0$ is about 51.0 ms, which is very close to the change period of 54.0 ms of the contact angle.

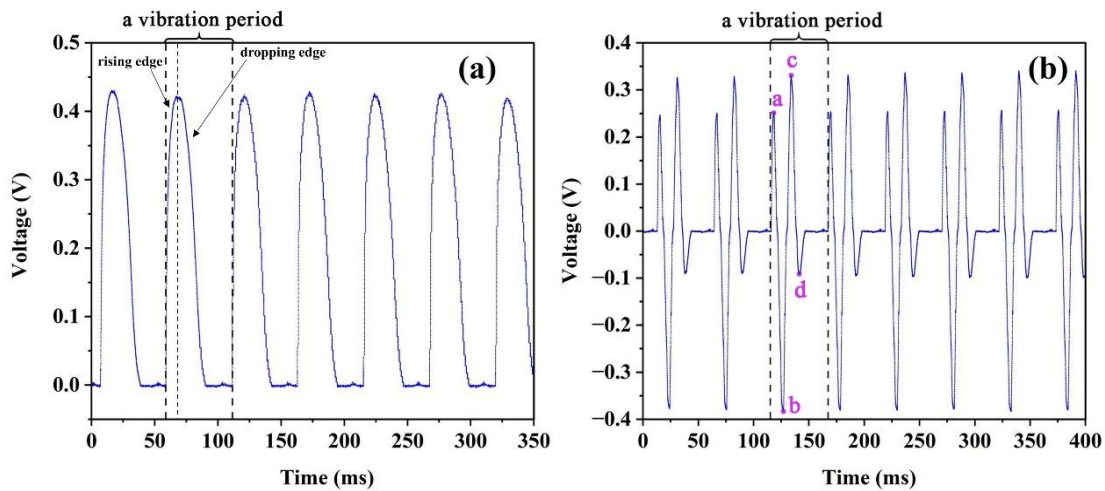

**FIG. 4. Voltage waveforms across sampling resistor (a) at DC voltages and (b) at AC voltages**

In addition, the voltage signal duration of the sampling resistor $R_0$ is 31.0 ms (rising edge + dropping edge), which is very close to the contact time of the droplet with the needle tip electrode of 28.0 ms obtained in Fig. 3(a). From Fig. 4(a), it can be seen that the voltage signal peak of the sampling resistor $R_0$ is asymmetric. Once the droplet is in contact with the needle tip electrode, the droplet spreads outward, however, the upper end of the droplet is always in contact with the needle tip electrode and the cross-sectional area of the liquid column continues to decrease, as seen from point 5 to point 8 in Fig. 3(b). This makes the resistance of the droplet continue to increase, resulting in the voltage of the sampling resistor $R_0$ constantly decreasing. Due to the strong inhomogeneous electric field near the needle tip electrode, the part of the droplet connected to the tip electrode is strongly polarized and a strong electric field force is exerted on it. Therefore, the liquid column will not easily leave the tip after touching it, resulting in a relatively slow drop in the voltage signal on the $R_0$. In contrast, the contact area expands rapidly as the droplet rebounds upward to touch the needle tip electrode, so the rising edge of the voltage signal is steeper than the dropping edge. Figure 4(b) represents the voltage waveform of the sampling resistor $R_0$ under AC (50Hz, 200 V) with time. Four peaks of the voltage signal appear in one vibration cycle at point a (3.0 ms), point b (11.4 ms), point c (18.6 ms), point d (26.7 ms). The period of the voltage signal on the sampling resistor $R_0$ is 50.0 ms, which is very close to the period of the change of contact angle, 46.0 ms. As seen in Fig. 3(c), there are three minimal values of the contact angle when the droplet is in contact with the needle tip electrode at point 5 (4.0 ms), 8 (12.0 ms), and 10 (19.0 ms), whose time are very close to the electrical signals of Fig. 4(b) at points a, b and c. The voltage at point d is too small to cause a significant change in droplet contact angle. The voltage on the sampling resistor $R_0$ is used as the sampling signal, and its magnitude is positively correlated with the voltage on the droplet. Therefore, we can conclude that the contact angle varies synchronously with the voltage signal of the droplet: the higher the voltage, the smaller the contact angle. More detailed analysis about the relationship between power supply voltage and the voltage on the sampling resistor is shown in Appendix 2.

### 3.1.3 Contact angle and amplitude vary with voltage

The large amplitude vibration of the droplet in the vertical direction is triggered by a combination of EWOD and surface tension effects. In the ordinary EWOD equilibrium state, the droplet's contact angle gradually saturates when the driving voltage is higher than a certain threshold.[29,30] In the case of PDMS film of 8.3 μm thickness (Appendix 3), droplet of 3 μL, the variation of contact angle (mean value) with DC voltage is shown in Fig. 5(a). The droplet contact angle variation can be divided into three stages: in stage I, when the voltage U < 80 V, cosine of the contact angle is linearly related to the square of the applied voltage. In stage II, when the voltage is 80 V < U < 120 V, as the applied voltage increases, the cosine of contact angle and the square of the applied voltage gradually deviate from the linear change, and its slope decreases with the increase of voltage. In stage III, when the voltage is 120 V < U < 210 V, the contact angle decreases from about 50° to 45° as the applied voltage increases.

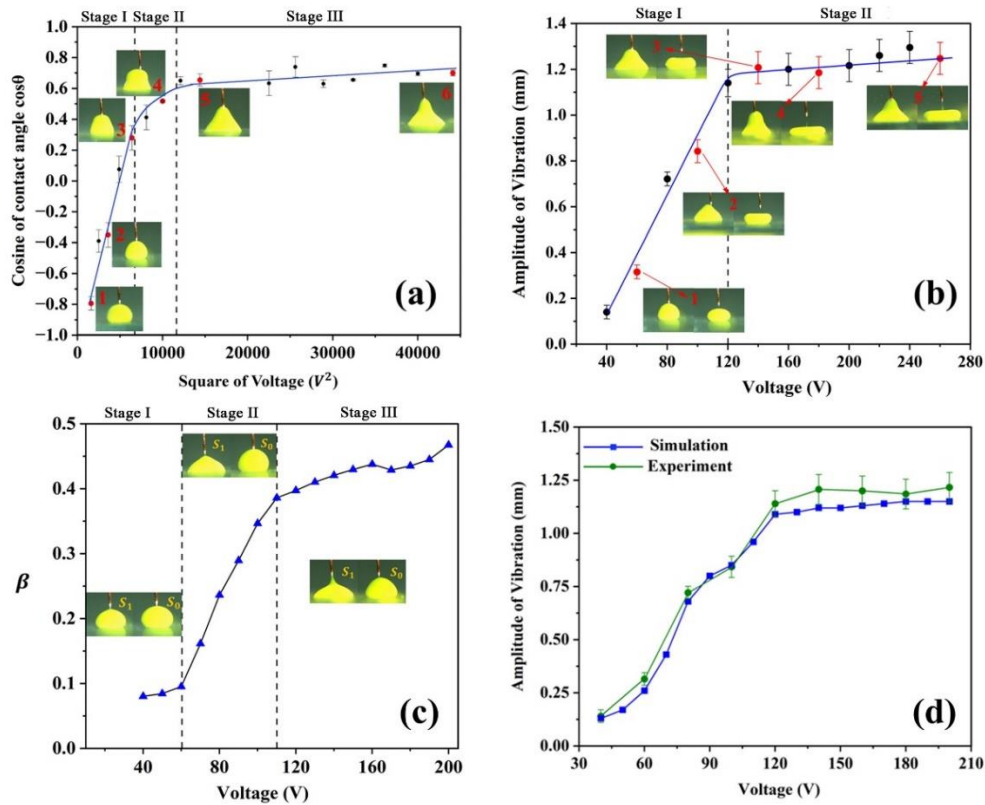

**FIG. 5. (a) Variation of contact angle of droplets with voltage during large vertical vibration; (b) Variation of droplet amplitude with voltage during large vertical vibration; (c) Dependence**

of the relative change in droplet surface energy with driving voltage; (d) Simulated versus experimental droplet morphology over time**

The results show that the contact angle saturation of the droplet still exist during the large vertical vibration. In the non-equilibrium vibration state of EWOD, the change of contact angle with applied voltage is similar to the equilibrium state change of EWOD. At the same time, the relationship between the vibration amplitude (the height difference between the highest and lowest points at the top of the droplet during the vibration) of the droplet and the applied voltage in the large vertical vibration state was also investigated. The DC voltage is between 40~260 V, and the amplitude of the droplet changes with the applied voltage, as shown in Fig. 5(b). When the voltage is 40 V<U<120 V, the droplet amplitude increases linearly with increasing voltage (stage I). When the voltage U> 120 V, the amplitude of the droplet slowly increases with the increase of voltage (stage II). It is noteworthy that the turning point of the droplet amplitude trend (~120 V) coincides with the contact angle saturation point (~120 V). This correlation is next explored by examining the change in surface energy of the droplet.

The spreading of liquid droplets under EWOD results in an increase in their surface energy (due to the change of the surface tension and the surface area), which is released when the droplet is out of contact with the needle tip electrode and further converted into kinetic energy of droplet motion, gravitational potential energy and frictional energy loss within the fluid.[31] We investigate the relationship between the surface energy and the driving voltage during large vertical vibration of the droplet with DC voltages ranging from 40 to 200 V. The coordinates of the positions of different points on the droplet profile are extracted using tracker and a polynomial function $x = f(y)$ is used to fit the droplet profile. Considering a droplet as an object with rotational symmetry around an axis, its lateral area is calculated by the following equation:

$$S = \int_0^{y_{max}} 2\pi f(y)\sqrt{1+f'(y)^2}\,dy \tag{8}$$

Where, $y_{max}$ is the droplet apex height. Calculate the lateral area $S_0 = \pi R_0^2$ of the droplet at the instant before contacting the needle tip electrode and the lateral area

$S_1 = \pi R_1^2$ at the instant before detachment. Then the change in surface energy of the droplet before and after contacting the needle tip electrode is:[27]

$$\Delta E_S = \gamma_{LL'}(S_1 - S_0) + (\gamma_{SL} - \gamma_{SL'})\pi(R_1^2 - R_0^2) \tag{9}$$

Where, $\gamma_{SL'}$ is the surface tension coefficient between the droplet and the tetrapropyloxy silane environment, $\gamma_{SL}$ is the surface tension coefficient between the droplet and the PDMS film and $\gamma_{SL'}$ is the surface tension coefficient between the tetrapropyloxy silane and the PDMS film (see Appendix 4) Let the relative change of surface energy be

$$\beta = \frac{\gamma_{LL'}(S_1 - S_0) + (\gamma_{SL} - \gamma_{SL'})\pi(R_1^2 - R_0^2)}{\gamma_{LL'}S_0 + (\gamma_{SL} - \gamma_{SL'})\pi R_0^2} \tag{10}$$

The dependence of the relative change in surface energy $\beta$ on the driving voltage is shown in Fig. 5(c). As seen from the picture, in stage I, the voltage is 40 V<U<60 V and the contact angle of the droplet is obtuse, the droplet morphology is ellipsoidal. There is only a small amount of deformation, and the relative change of the surface energy increases slowly with the increase of the driving voltage. In stage II, the voltage is 60 V<U<110 V, the contact angle of the droplet is acute, the droplet morphology is ellipsoidal ($S_0$) and conical ($S_1$), and the relative change of surface energy increases rapidly with the increase of driving voltage. In stage III, with voltage 110 V<U<200 V, the morphology of the droplet no longer changes significantly after contact with the needle tip electrode due to the more pronounced contact angle saturation effect. Therefore, the relative change in surface energy increases slowly with the increase in driving voltage. When droplet contacts the needle tip electrode, the energy of the electric field is converted into the surface energy of the droplet. Because during the free rebound of the droplet the surface energy, kinetic energy, and potential energy converts mutually, so a larger droplet surface energy change leads to a larger vibration energy and amplitude. A comprehensive analysis shows that before the contact angle saturation, the higher the driving voltage, the smaller the droplet contact angle, and the flatter the droplet becomes; Therefore, the more surface energy it stores, and then the bigger the amplitude. After the contact angle saturation, the contact angle saturation effect suppresses the change of droplet morphology, the vibrational energy of the droplet increases slowly, and the amplitude of the droplet also increases slowly. Therefore, the saturation of the contact angle in contact with the needle tip electrode necessarily implies saturation of the droplet amplitude.

**3.1.4 Numerical simulation of the vibration**

The evolution of droplet morphology was simulated according to the equations (2) to (4) during large amplitude vibration. The initial shape of the droplet was selected to be elliptical in cross-section, a = 1.05 mm, b = 0.95 mm, z = 1.45 mm, H = 1.4 mm, and $\theta_0$ =120º, $\theta_1$ =45º (which corresponds to the voltage of 200 V, look at Fig.5(a) for reference). The simulated droplet morphology and the experimental droplet morphology over time are shown in Fig. 6. From the simulation results, it can be seen that during 0~4 ms: the droplet is free to rebound and does not touch the tip of the needle. During 8~24 ms: after the top height reaches H, the droplet touches the needle tip electrode and the contact angle of the droplet changes. Under the effect of inertia, the top of the droplet continues to move upward for a small distance, after which the droplet evolves into a cone shape and begins to fall. During 28~40 ms: after the height of the top of the droplet is less than H, the droplet is detached from the needle tip electrode and the contact angle is immediately restored, gradually evolving from a raised shape to a flattened and then concave shape. During 40-52 ms: the droplet rebounds upwards from a depressed shape and gradually evolves into a flattened and then raised shape. At the moment of 28 ms: the top of the experimental droplet is still attached to the needle tip electrode, but the simulated droplet has detached from the tip, which may be caused by the strong electric field near the needle tip electrode that has a strong attraction to the droplet. Overall, the droplet morphology evolutions from the theoretical simulations are almost identical to those from the experiments.

In the process of large vertical vibration, the droplet amplitude changes significantly with the change of driving voltage. Further theoretical simulation of the effect of driving voltage on droplet amplitude has been constructed. The initial droplet shape parameters are: a=0.85 mm, b=0.75 mm, z=1.4 mm, H=1.3 mm, contact angle $\theta_0$ =145°. The DC voltage is 40 to 200 V (The contact angle $\theta_1$ was chosen as seen in Appendix 5). The simulated droplet amplitude varies with voltage as shown in Fig. 5(d). Between 40 and 120 V, the simulated droplet amplitude increases significantly with increasing voltage; between 120 and 200 V, the simulated droplet amplitude increases slowly with increasing voltage. The simulated amplitude variation is almost the same trend as the experimental amplitude variation, which also indicates that the contact angle saturation effect has a decisive influence on the trend of droplet amplitude variation.

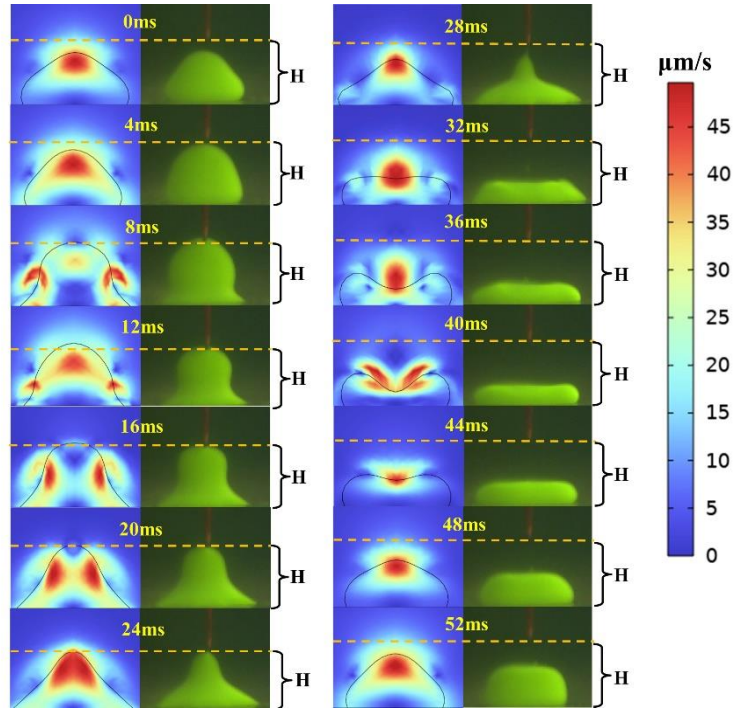

**FIG. 6. Comparison of simulation and experimental results of droplet amplitude variation with voltage**

## 3.2 Horizontal motion

### 3.2.1 Conductivity of the PDMS film

The PDMS film acts as a superhydrophobic coating which greatly reduces the resistance to horizontal movement of the droplet. However, in ESI experiments, it remains to be verified that the droplet can still be considered as grounded when the PDMS film separates the droplet from the substrate. We first analyze the electrical conductivity of this film. The equivalent circuit of the system is shown in the Fig.7(a).

In the circuit, the capacitance and resistance of the PDMS substrate can be estimated using $C_2 = \dfrac{\varepsilon S}{h}$ and $R_2 = \rho_2 \dfrac{h}{S}$, respectively; The droplet resistance can be taken as $R_1 = \rho_1 \dfrac{2b}{\pi a^2}$. Where $\varepsilon = 2.662 \times 10^{-10}$ F/m is the dielectric constant of PDMS, $h$=3.5 μm is the thickness of the PDMS film, $\rho_1 = 1.5 \times 10^8$ Ω·m and $\rho_2 = 2.9 \times 10^{12}$ Ω·m is the resistivity of the droplet and PDMS, respectively, $S \approx 1$ mm$^2$ is the base area of the droplet. To calculate the capacitance $C_2$ between the droplet and the needle tip electrode, the voltages from 400 to 1000 V were chosen to simulate the charge of the droplet at different horizontal distances ($d$) from the needle tip electrode to the droplet when the droplet is grounded, and the result is shown in the Fig.7(b). The charge of the droplet is on the order of $10^{-10}$ C, and the higher the driving voltage, the closer the droplet is to the needle tip electrode, and the larger the charge capacity of the droplet is. The upper limit of the capacitance $C_2$ can be calculated using the charge capacity result under the driving voltage of $U = 1000$ V and the distance of $d$=0 mm: $C_2 = \dfrac{Q}{U} = 0.3$ pF. The system of differential equations satisfied by the equivalent circuit is

$$\begin{cases} C_1 \dfrac{d}{dt}(U_1 - U) = \dfrac{U_2 - U_1}{R_1} \\ C_2 \dfrac{dU_2}{dt} = \dfrac{U_1 - U_2}{R_1} - \dfrac{U_2}{R_2} \end{cases} \quad (11)$$

The right upside of Fig.7(a) is the numerical result of the system of differential equations under $U = 1000$ V. The calculation results show that after the power is turned on, the droplet reaches the equipotential ($U_1 = U_2 = 1000$ V) at about 0.74 s, and then the droplet potential slowly decreases. Since the droplet reaches the equipotential quickly and the droplet potential is much smaller than the applied voltage after the equipotential, we can approximate that the droplet is always grounded during the experiment. In the later experiment, we approximate that at the moment of turning on the voltage, the droplet is already charged under ESI and reaches electrostatic equilibrium.

### 3.2.2 Experimental and numerical analysis of the horizontal motions

Under the action of the electric field force, the droplet accelerates towards the needle tip electrode. During this movement, the droplet is subjected to electrostatic forces $F_x$ and $F_z$, environmental resistance $f$, droplet gravity $mg$, buoyancy $f_b$, and the normal force $N$ of the substrate to the droplet, as shown in the Fig.7(c). Assuming that the environmental resistance $f$ is the sum of the frictional resistance of the substrate to the droplet and the viscous resistance of the liquid environment to the droplet, the frictional resistance of the substrate to the droplet is proportional to the normal force $N$ of the substrate to the droplet, and its proportional coefficient is $\mu$. The viscous resistance of the liquid environment to the droplet is proportional to the scale of the droplet (semi-major axis a) and the velocity $\upsilon$ of the droplet, and the proportional coefficient is $k$.

$$f = \mu N + k \cdot a \cdot \upsilon = \mu(mg - F_z - f_b) + k \cdot a \cdot \upsilon \quad (12)$$

where

$$f_b = \rho' V g \quad (13)$$

$$mg = \rho V g \quad (14)$$

$\rho$ is the droplet density and $\rho'$ is the density of the surrounding liquid. Therefore, under the joint action of electric field force and environmental resistance, the kinetic equation of droplet motion is:

$$m\frac{d\upsilon}{dt} = F_x - \mu\left[(\rho-\rho')Vg - F_z\right] - k \cdot a \cdot \upsilon \quad (15)$$

where $m$ is the droplet mass.

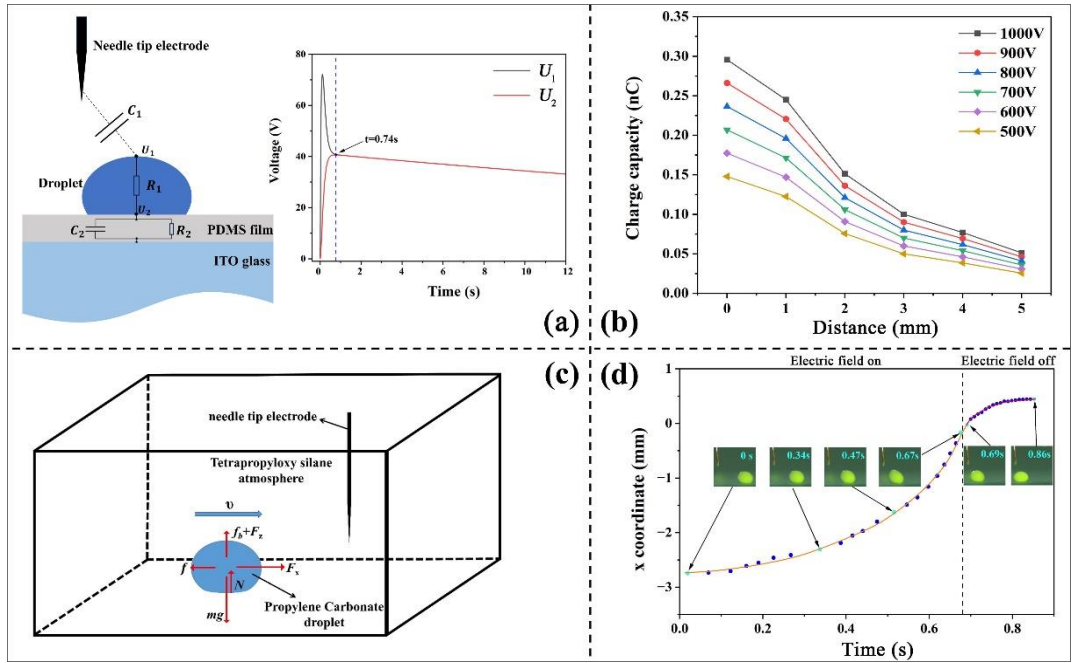

FIG. 7. (a) The equivalent circuit of the system (b)The relationship between the charge of the droplet and the distance between the needle tip and the electrode under different voltages; (c) Force analysis of droplet; (d) During the process of acceleration and deceleration, the relationship between the coordinate of the droplet center position and the time

To measure the resistance encountered during the movement of the droplet, the driving voltage was adjusted to 1000 V, the semi-major axis of the droplet was 0.8 mm during the experiment, and there was a certain distance between the droplet and the needle tip electrode at the initial moment. Under the action of the electric field force, the droplet accelerates to move towards the needle tip electrode. After the droplet has a certain speed, remove the electric field, and the droplet decelerates under the action of friction and viscous resistance. The droplet morphology and its center position coordinate $x$ versus time are shown in Fig.7(d). It can be seen from Fig.7(d) that when $t$ <0.67 s, under the action of the electric field force, the droplet accelerates towards the needle tip electrode and deforms, and the closer the droplet is to the tip electrode, the greater the deformation. When $t$ >0.67 s, remove the electric field, the deformation of the droplet recovers quickly, and the droplet decelerates until it stops. Let $F_x = F_z = 0$ in the equation (15), then the functional relationship of the coordinate $x$ of the center of the droplet with time $t$ can be obtained as follows:

$$x(t) = -\left(\frac{m\mu(\rho-p')Vg}{k^2a^2} + \frac{mv_0}{k^2a^2}\right)\left(e^{-\frac{ka}{m}t} - 1\right) - \frac{\mu(\rho-p')Vg}{ka}t + x_0 \quad (16)$$

where $v_0$ and $x_0$ are respectively the velocity and coordinates of the center of the droplet when the electric field is removed. It can be seen from Fig.7(d) that when $t$ =0.69 s, the deformation of the droplet has basically recovered; When 0.69 s<$t$<0.86 s, the $x-t$ data points of the droplet are fitted in the form of $x = -A(e^{-Bt}-1) - Ct$ function. From the fitting results, it can be known that the values of parameters $k$ and $\mu$ are 0.02782 kg/(m·s) and 0.00818, respectively.

In order to obtain the force on the droplet during the acceleration process, we calculated the electric field forces $F_x$ and $F_z$ on the droplet using the equations (6) and (7). The forces at different horizontal distances from the needle tip electrode at 400 V and 1000 V are shown in Fig.8(a-d). When the voltage is 400 V, the electric field force $F_x$ on the droplet first increases and then decreases with the decrease of the distance, and decreases to zero right below the needle tip. The electric field force $F_z$ on the droplet increases monotonously as the distance decreases. At the voltage of 1000 V, the changing trend of the electric field force $F_x$ and $F_z$ on the droplet is the same as that at 400 V. A polynomial fitting is performed on the simulated data points of $F_x \sim d$ and $F_z \sim d$ respectively to obtain the continuous function $\tilde{F}_x(d)$ and $\tilde{F}_z(d)$ of the electric field force $F_x$ and $F_z$ on the droplet.

Considering electric field force and resistance together, the continuous functions $\tilde{F}_x(d)$ and $\tilde{F}_z(d)$, and the parameters $k$ and $\mu$ are substituted into equation (15) for numerical solution, which gives the theoretically predicted curves of the horizontal distance between the droplet center and the tip of the needle as a function of time, as shown in Fig.8(e, f) (solid line). The experimentally measured data of distance over time are shown in Fig.8(e, f) (scattered dots). As can be seen in Fig.8(e, f), at 400 V, the droplet first accelerates and then decelerates gradually, and then stops moving right below the needle tip electrode, with no obvious deformation of the droplet and no contact with the needle tip electrode. At 1000 V, the droplet is always accelerating and is significantly deformed near the needle tip electrode, at which time the droplet is strongly attracted by the needle tip electrode before it touches the needle tip electrode, which causes the droplet to vibrate substantially (see video II, "droplet accelerates at 1000 V"). The theoretically predicted distance versus time is in good agreement with the experimental results. It is worth noting that since the droplet gravity minus the buoyant is

$$F_{G} = (\rho - \rho')gV \approx 5.6\ \mu N \tag{18}$$

As can be seen from Fig.8(d), when the droplet is close to the needle tip electrode (d<0.75 mm), there is $\tilde{F}_z(d) > F_G$, thus the droplet is attracted upwards and touches the needle tip electrode, which is also in good agreement between theory and experiment.

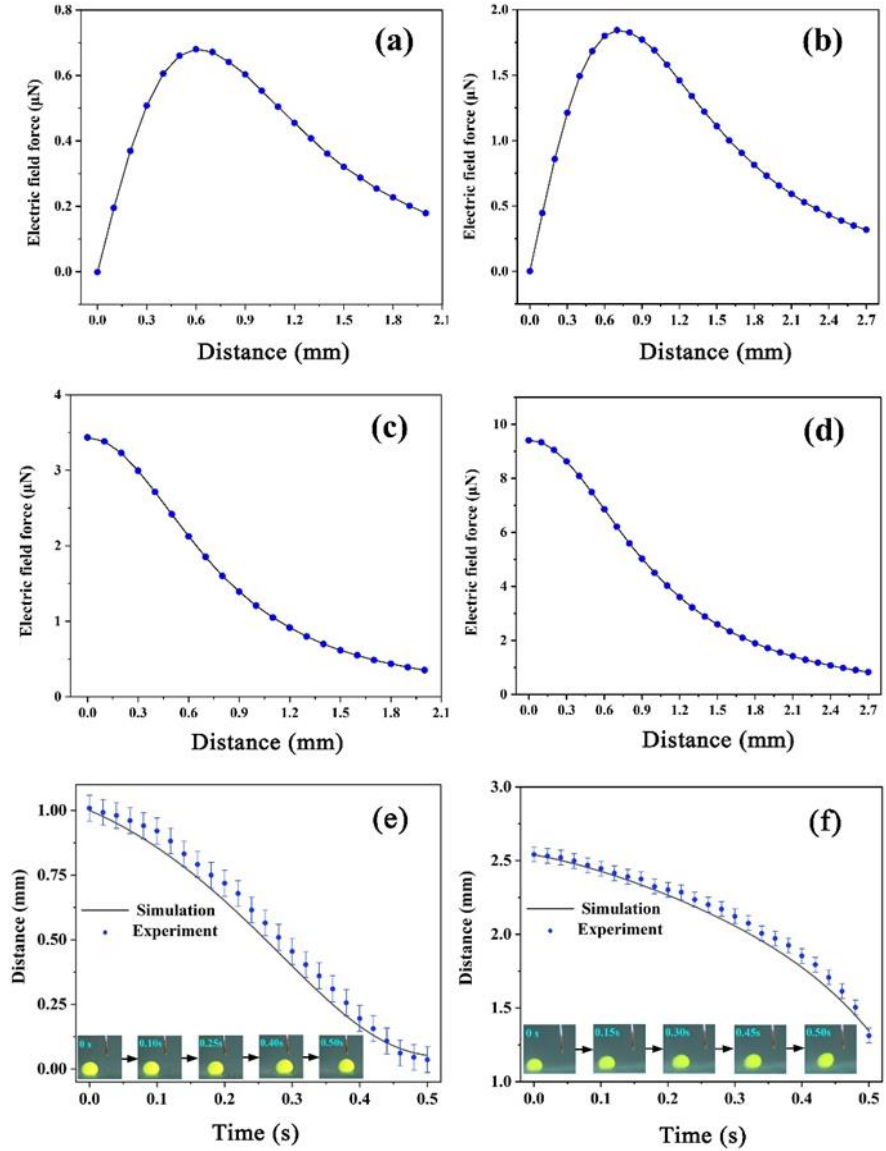

FIG. 8. Simulation analysis of force on droplet (a) at 400 V, the electric field force $F_x$ on the droplet varies with distance; (b) at 1000 V, the electric field force $F_x$ on the droplet varies with distance; (c) at 400 V, the electric field force $F_z$ on the droplet varies with distance; (d) at 1000 V, the electric field force $F_z$ on the droplet varies with distance. Experimental data and theoretical simulation curves for droplet motion (e) at 400 V; (f) at 1000 V

## 4.Conclusion

In this paper, tetrapropoxysilane is used as an immersing environment to study the electric-field induced motion of a droplet on PDMS films. In the large vertical vibration caused by EWOD, a voltage of 200 V was applied to the needle tip electrode and the

PDMS-ITO substrate. The droplet appeared to vibrate steadily and substantially with a vibration period of about 60 ms at 1.32 mm < H < 2.04mm. When H < 1.32 mm, the droplet enters the EWOD equilibrium state and the vibration disappears. In the vibration under a DC voltage of 200 V, the variation of droplet morphology and contact angle with time shows that when the droplet contacts and detaches from the needle tip electrode, the contact angle is about 40º and 130º, respectively. The contact time of the droplet with the needle tip electrode, 28 ms, is very close to the duration of the electrical signal in one cycle, 31 ms. Under AC, when a droplet contacts the needle tip electrode, the contact angle of the droplet varies with the AC signal, and the peak moments of the AC signal (3.0 ms, 11.4 ms, 18.6 ms) coincide with the moments of the minimum contact angle (4.0 ms, 12.0 ms, 19.0 ms). When the voltage U < 120 V, the amplitude of the droplet increases significantly with increasing voltage and the contact angle decreases significantly with increasing voltage. At voltage U > 120 V, the droplet amplitude and contact angle saturate with increasing voltage. Simulations show the decisive role of the contact angle saturation effect on the trend of the droplet amplitude. In the horizontal motion triggered by ESI, it is verified that the droplet can be regarded grounded during the experiment and the charge of the droplet is about $10^{-10}$ C order of magnitude. By studying the deceleration process of the droplet, the frictional resistance coefficient of the droplet was experimentally measured to be about 0.00818 and the viscous resistance coefficient to be about 0.02782 kg/(m·s). The theoretical simulation shows that the electric field force on the droplet is on the order of $10^{-6}$ N. The electric field force $F_x$ increases and then decreases with the distance, and it decreases to zero directly below the tip of the needle. The electric field force $F_z$ on the droplet increases monotonically with distance. The theoretical calculation of the horizontal acceleration of the droplet at 400 V and 1000 V has been constructed. At 400 V, the droplet first accelerates and then decelerates, and comes to a standstill directly under the needle tip electrode; At 1000 V, the droplet is strongly attracted by the needle tip electrode and accelerates, and finally touches the needle tip electrode.

## 5. Conflicts of interest

The authors declare no competing financial interest.

## 6. Acknowledgements

This work was financially supported by the Teaching Research Project of University of Science and Technology of China 2023xxskc06 and 2022xjyxm021; Virtual Simulation Experiment Teaching Project of Anhui 2022xnjys029; and the Provincial Teaching Research Project of Anhui 2021jyxm1717.

## 7. Data availability

The data that support the findings of this study are available from the corresponding author upon reasonable request.

## 8.REFERENCE

**Appendix information**

**1. The experimental methods were adopted as follows:**

(1) PDMS film was deposited on the surface of ITO conductive glass.

The polydimethylsiloxane reagent (containing the main agent and curing agent) of Sylgard184 was used to prepare the polydimethylsiloxane prepolymer (PDMS), and its mass ratio was the main agent: curing agent: cyclohexane = 10:1:2. The PDMS's dielectric constant of Sylgard184 is approximately 25.1 pF/m. The configured PDMS prepolymer is placed in a vacuum drying chamber for 15 min to remove tiny air bubbles from the PDMS. Cover the corner of the clean ITO glass with capton tape for the ITO substrate connection electrode. Place the ITO glass on the homogenizer suction cup and fix it by an air pump. Pipette approximately 0.5 mL of PDMS prepolymer and evenly coat the ITO glass. Preset the maximum speed of spin coating and turn on the homogenizer. The rotational speed is set sequentially: 35 seconds to 3000 r/min, 50 seconds to a maximum speed of 5000 r/min (or 7500 r/min), maintain the maximum speed for 60 seconds, slow down to 3000 r/min in 45 seconds, stop rotation in 30 seconds. The above-mentioned coated ITO glass can be coated with a PDMS film (dielectric layer) of a certain thickness by baking at 70 °C for 15 minutes.

(2) The sample cell and droplet manipulation environment

The sample cell was made of white glass bonded together, ITO glass with PDMS film spin-coated was placed on the iron base at the bottom of the cell, and the energizing electrode was a column magnet, which took into account the roles of energizing and fixing the substrate. The droplets used in this experiment were propylene carbonate (99%, Aladdin), and a certain amount of the colorant rose red (18NR-T, Beijing Jinge Co. Ltd.) was added to the droplets for easy observation. In order to provide a low resistance environment for the droplet motion, both the droplet and the ITO glass spin-coated with PDMS were completely submerged in tetrapropoxysilane (97%, Aladdin).

(3) Photographing and analyzing droplet morphology

The movement of the droplets was illuminated and photographed using a common

LED light source with a high-speed video camera (FASTCAM SA-X2, Photron Limited), and the video frame rate was adjusted to 1000 fps. The video was imported into tracker software for frame-by-frame analysis to make quantitative measurements of the top height during droplet vibration, changes in contact angle, and changes in the position of the droplet during attraction by the needle tip electrode.

(4) The construction of the circuit

In this experiment, a DC voltage (CE1000001T, Rainworm) and a 50 Hz AC voltage (TDGC2-KVA, Shanghai Langge Co. Ltd.) were used to manipulate the droplet movement. One pole of the power supply was connected to the ITO glass with a PDMS film spin-coated, and the other pole was connected to a needle-tip electrode through the sampling resistor $R_0$. The voltage waveforms at both ends of sampling $R_0$ were measured by an oscilloscope (Aliglent Technologies InfiniiVison DSO-X 3052A). To improve the signal-to-noise ratio of the measured data, the sampling R_0 was selected as 1 kΩ in the experiment.

## 2. The relationship between power supply voltage and the voltage on the sampling resistor

In this experiment, the thickness of PDMS is about 3.5 μm and its dielectric constant $\varepsilon$ is about 25.1 pF/m. The radius of the bottom surface of the droplet is about 1 mm, and an equivalent capacitance (~20 pF) exists between the droplet and the PDMS. Therefore, the time constant $\tau$ ($R_0C$) of the whole circuit is about a few tens of nanoseconds, which has a negligible effect on the circuit. The droplet and the PDMS film can be regarded as a pure resistance ($R$), and the leakage capability between the droplet and the PDMS film is reflected by measuring the voltage signal of the sampling resistance $R_0$. As the DC voltage increases from 40 to 260 V, the large vertical vibration of the droplet is induced, and the relationship between the peak voltage of the sampling resistor $R_0$ and the applied voltage is shown in Fig. S1. The resistance $R$ between droplet and the PDMS decreases with applied voltage and has a value of about $2.5\times10^6$ Ω at 40 V and $3.4\times10^5$ Ω at 260 V.

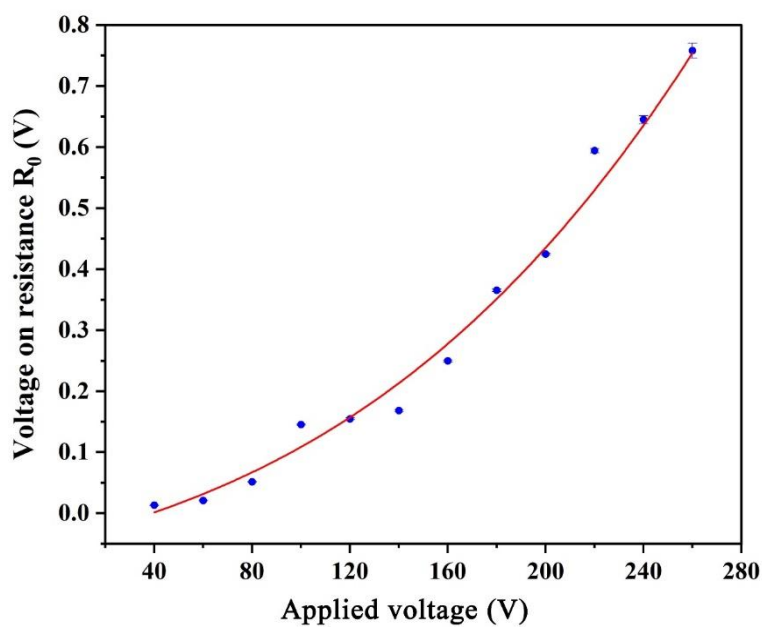

Fig.S1 Variation of peak voltage of sampling resistor $R_0$ with applied voltage

## 3. The relationship between PDMS film thickness and rotation speed

The maximum rotational speed of spin coating was set at 5000 r/min and 7500 r/min, respectively, and a scanning electron microscope (SU8010, Hitachi) was used to observe the cross-section of the PDMS film to obtain the thickness of the spin-coated PDMD film at different rotational speeds, as shown in Table S1.

Table S1 Relationship between rotational speed and PDMS film thickness

| Rotational speed (r/min) | Average film thickness (μm) | Representative Figure |
|---|---|---|
| 5000 | 8.3 | (SEM image, 8.37 μm) |
| 7500 | 3.5 | 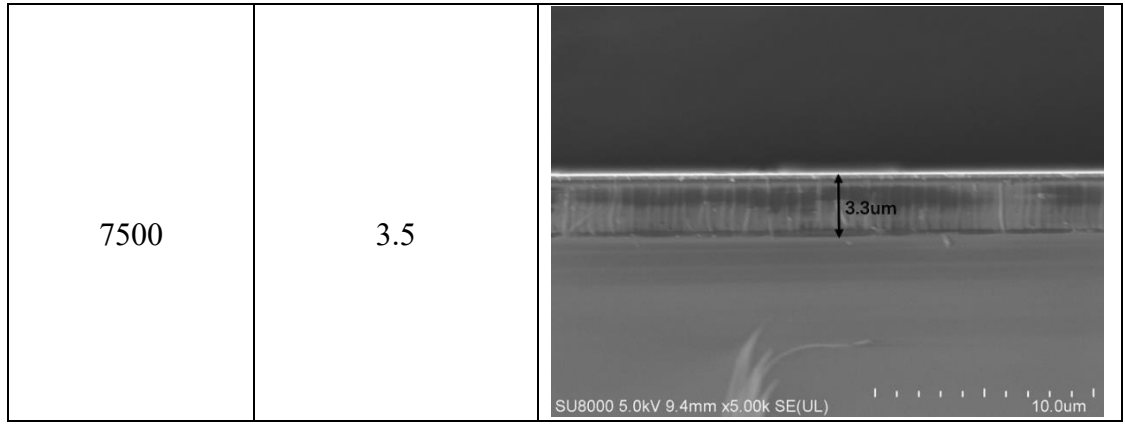 |

## 4. Estimation of surface tension coefficient

(1) $\gamma_{LL'}$ between the droplet and the tetrapropoxysilane environment

A PDMS substrate with a spin-coating speed of 7500 r/min was selected, the DC voltage was controlled from 0 to 31 V. The needle tip electrode was inserted into the droplet and the droplet was brought to the EWOD equilibrium state, and the relationship between the cosine of the contact angle of the droplet and the square of the voltage is shown in Fig. S2(a). According to the Lippmann-Young equation, the relationship between the cosine of the contact angle and the square of the voltage should be linear. Because of the saturation effect of the contact angle, the higher the voltage, the more the relationship deviates from linearity. Therefore, the first six data points of Fig. S2(a) were selected for linear fitting, as shown in Fig. S2(b), with a fitting slope k = 0.00286. As shown in Table S1, the thickness of the PDMS film is about 3.5 μm, and the surface tension coefficient between the droplet and the surrounding environment is calculated by substituting into the Lippmann-Young equation to be 1.227 mN/m, which means that the surface tension coefficient is taken to be the same value when simulating the vibration process of the droplet.

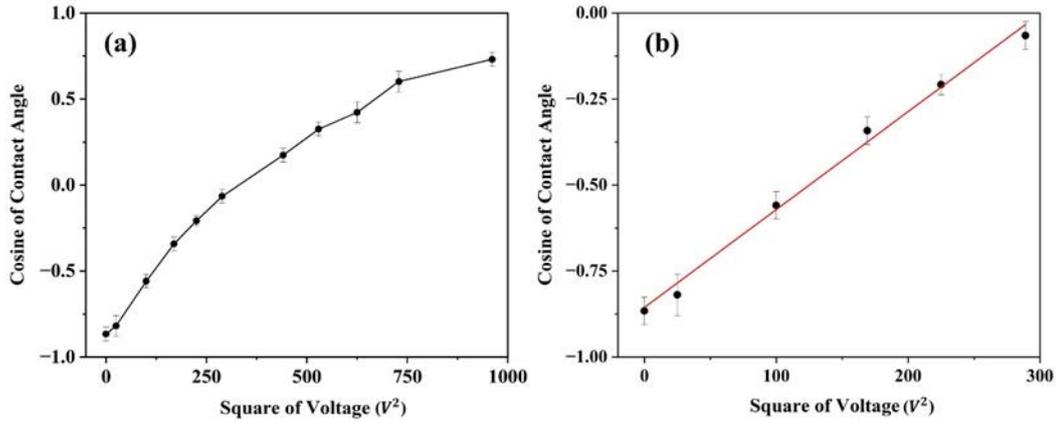

Fig. S2 Surface tension coefficients between droplets and tetrapropoxysilane environments

(2) Difference in surface tension coefficient $\gamma_{SL} - \gamma_{SL'}$

The droplet is placed statically on the substrate surface without the influence of an external electric field, according to the Young's equation for static equilibrium:

$$\gamma_{LL'} \cos\theta + \gamma_{SL} = \gamma_{SL'}$$

The droplet contact angle $\theta = 145.9°$ (i.e., Fig. S3 is the complementary angle to the contact angle) gives

$$\gamma_{SL} - \gamma_{SL'} = -\gamma_{LL'} \cos\theta = 1.017 \text{ mN/m}$$

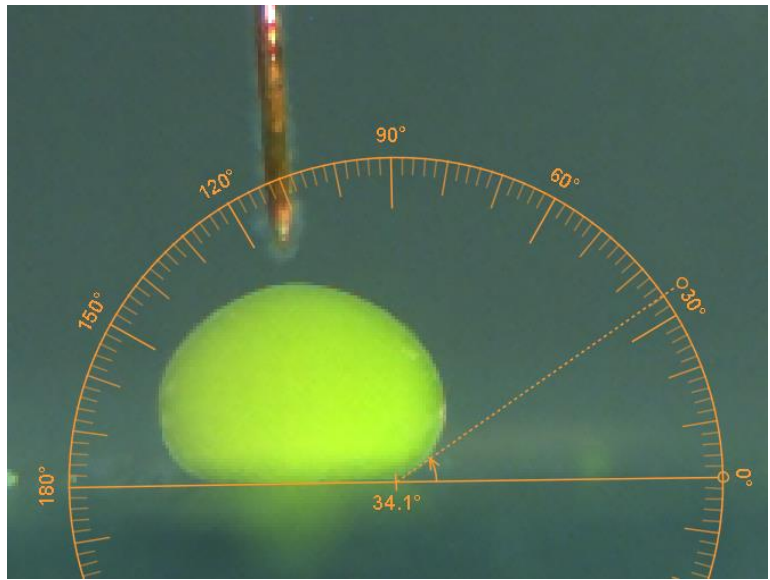

Figure S3 Contact angle measurement of a droplet in the absence of an external electric field

We make a complementary analysis of the functional relationship of the droplet in the presence of EWOD effects, giving a calculation of $\gamma_{SL} - \gamma_{SL'}$ from the energy point of view. The needle tip electrode was inserted into the droplet, and a DC voltage was applied between the needle tip electrode and the PDMS membrane to bring the droplet to the EWOD equilibrium state. In the quasi-static state, the voltage was increased from $U_1 = 13 \text{ V}$ to $U_2 = 21 \text{ V}$, and the relationship between the voltage and the radius of the bottom surface of the droplet was measured, as shown in Fig.S4. The red curve is the result of the third-degree polynomial fit $U(R)$.

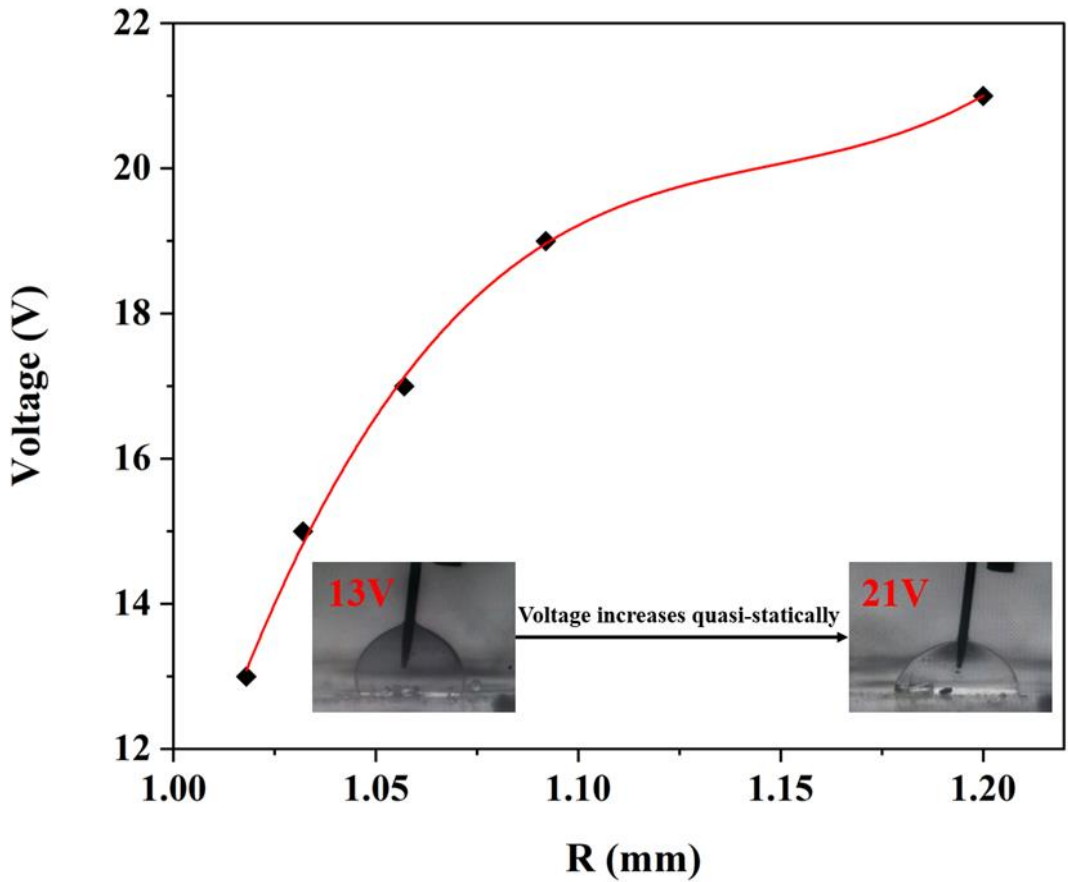

Fig. S4 DC voltage vs. radius of droplet bottom surface

When the droplet reaches the EWOD equilibrium state, the electric field force per unit length in the horizontal direction at the three-phase contact line between the droplet, the PDMS substrate, and the tetrapropoxysilane environment can be found out by using the electric field stress tensor to represent the electric field force and applying the conformal mapping to the edge region of the droplet contact angle [32]:

$$F_h = \frac{\varepsilon_d U^2}{2d}$$

Where $\varepsilon_d$ is the dielectric constant of the PDMS substrate, $U$ is the applied DC voltage, and $d$ is the PDMS film thickness. In the process of increasing the applied voltage from $U_1$ to $U_2$, the radius of the droplet bottom surface increases from $R_1$ to $R_2$, so the work done by the electric field force on the droplet in this process is

$$W_e = \int_{R_1}^{R_2} \int_0^{2\pi} \frac{\varepsilon_d U^2(R)}{2d} R dR d\theta$$

Substituting the data and integrating numerically gives $W_e = 1.894 \times 10^{-9}$ J. Calculate the droplet's lateral area as described in the main text to obtain the incremental increase in droplet lateral area as the voltage is increased from $U_1$ to $U_2$ is $\Delta S = 0.696 \times 10^{-6} \text{m}^2$. This in turn gives the surface energy increment on the contact surface between the droplet and the tetrapropoxysilane as

$$\Delta E_{LL'} = \gamma_{LL'} \Delta S = 1.004 \times 10^{-9} \text{ J}$$

As the voltage increases from $U_1$ to $U_2$, the droplet center of mass decreases, and according to the formula for the droplet center of mass height

$$y_c = \frac{\int_{y_1}^{y_2} y\pi x^2(y) dy}{\int_{y_1}^{y_2} \pi x^2(y) dy}$$

The increment of the gravitational potential energy of the droplet (taking into account the buoyancy force) can be obtained as

$$\Delta E_p = (\rho - \rho') g \Delta y_c \int_{y_1}^{y_2} \pi x^2(y) dy = 3.54 \times 10^{-10} \text{ J}$$

Therefore, the energy changes due to droplet spreading is

$$\Delta E_{SL} - \Delta E_{SL'} = W_e + \Delta E_P - \Delta E_{LL'} = 1.244 \times 10^{-9} \text{ J}$$

Neglecting the energy dissipation due to drag, the difference between the required surface tension coefficients is obtained as

$$\gamma_{SL} - \gamma_{SL'} = \frac{\Delta E_{SL} - \Delta E_{SL'}}{\pi(R_2^2 - R_1^2)} = 0.877 \text{ mN/m}$$

Therefore, the difference between the surface tension coefficients calculated directly from the Young equation is not much different from the results calculated from the energy conservation. In the main text, $\gamma_{SL} - \gamma_{SL'}$ is taken to be 1.017 mN/m.

## 5. Simulation results of droplet amplitude variation with voltage

According to Fig. 6(a) in the main text, the contact angles of the droplets at different voltages are read and taken as the equilibrium contact angle used in the simulation, and the simulation results of droplet vibration at different voltages are obtained. Simulated images of the highest and lowest positions reached at the top of the droplet at different voltages are shown in Table S2.

Table S2 Simulated images of the highest and lowest positions reached by the top of the droplet

| Voltage $\theta_1$ | Droplet vertex z-coordinate Image of the highest position of the droplet vertex | Droplet vertex z-coordinate Image of the highest position of the droplet vertex |
|---|---|---|
| 40 V $\theta_1=130°$ | 1.22 mm 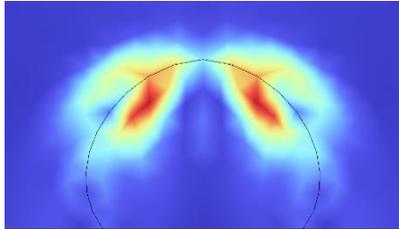 | 1.09 mm 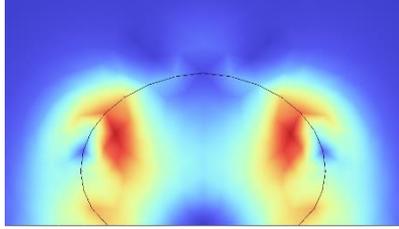 |
| 50 V $\theta_1=121°$ | 1.22mm 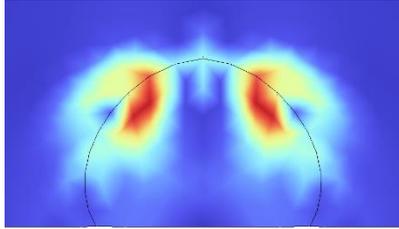 | 1.05mm 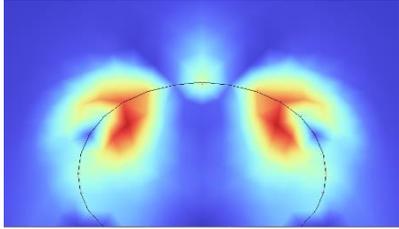 |
| 60 V $\theta_1=110°$ | 1.23mm 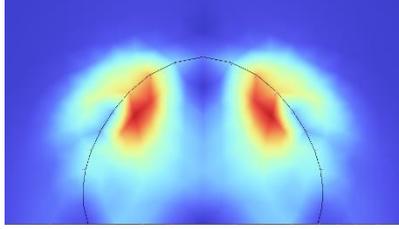 | 0.97mm 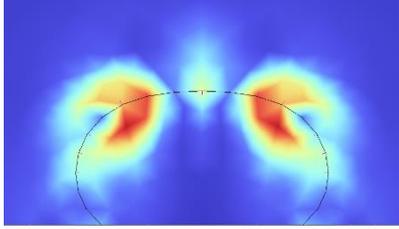 |

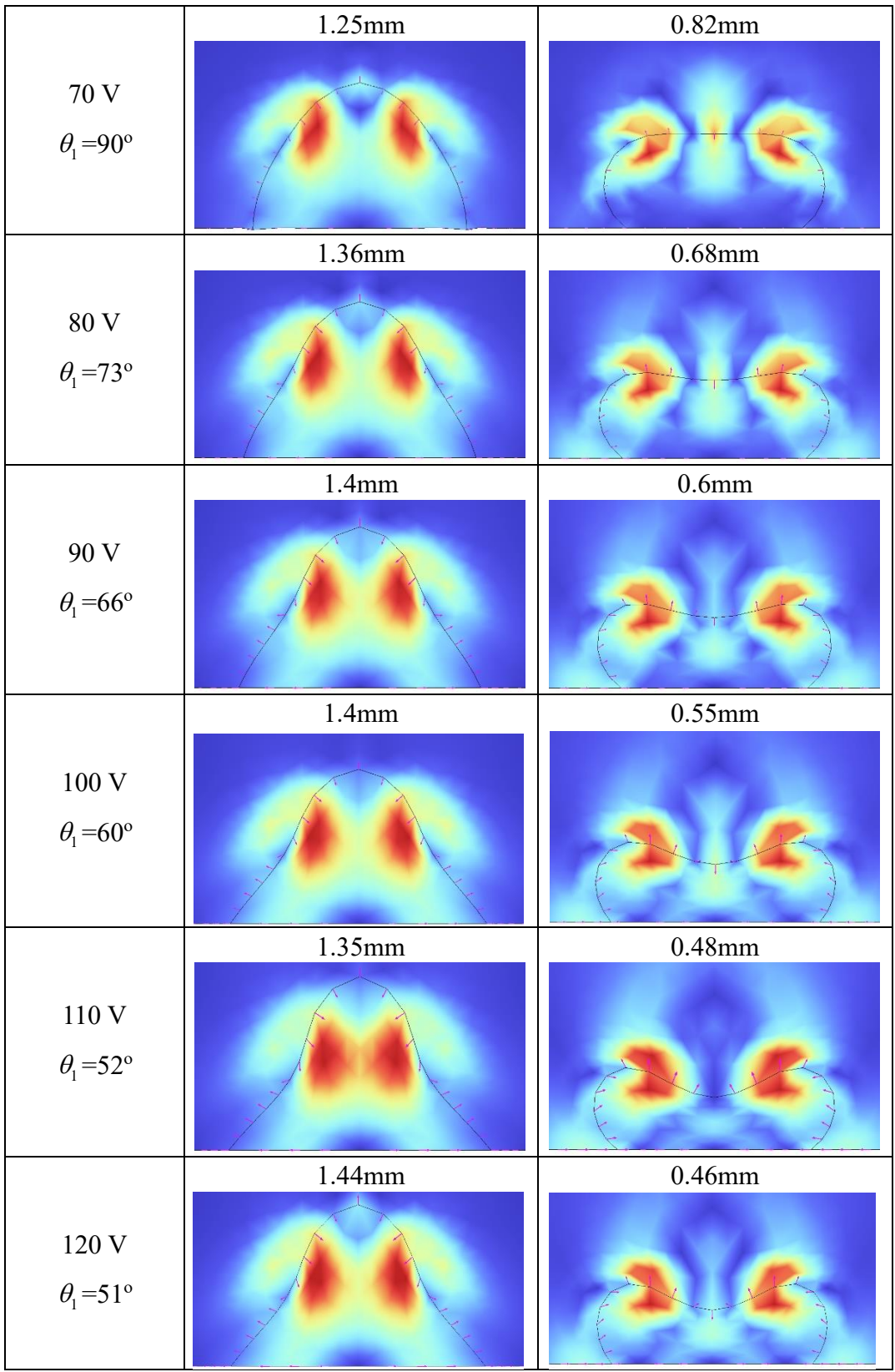

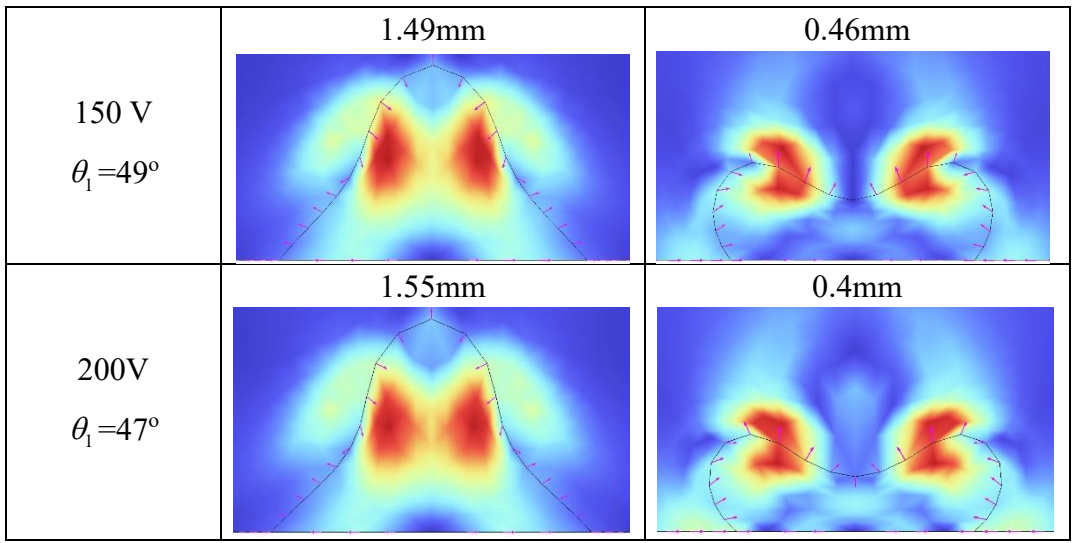